\documentclass[sigconf]{acmart}
\usepackage{listings}
\usepackage{xcolor}

\acmConference[SBQS26]{Brazilian Symposium on Software Quality}{Nov 10--13, 2026}{Maringá, PR}
\setcopyright{none}
\renewcommand\footnotetextcopyrightpermission[1]{}

\AtBeginDocument{%
  }

\begin{document}
\renewcommand\abstractname{ABSTRACT}
\renewcommand\keywordsname{KEYWORDS}
\renewcommand\refname{REFERENCES}
\hypersetup{hidelinks}

\title{Context Matters: Improving LLM-Based Code Summarization with Hierarchical Information}

\author{Alexandre Petrachini}
\orcid{0009-0003-8027-772X}
\affiliation{%
  \department{Instituto de Computação}
  \institution{Universidade Estadual de Campinas (UNICAMP)}
  \city{Campinas}
  \state{São Paulo}
  \country{Brazil}
}
\email{a212306@dac.unicamp.br}

\author{Breno Bernard Nicolau de França}
\orcid{0000-0002-4531-1473}
\affiliation{%
  \department{Instituto de Computação}
  \institution{Universidade Estadual de Campinas (UNICAMP)}
  \city{Campinas}
  \state{São Paulo}
  \country{Brazil}
}
\email{bfranca@unicamp.br}

\author{Guido Araújo}
\orcid{0000-0003-4869-5190}
\affiliation{%
  \department{Instituto de Computação}
  \institution{Universidade Estadual de Campinas (UNICAMP)}
  \city{Campinas}
  \state{São Paulo}
  \country{Brazil}
}
\email{guido@ic.unicamp.br}

\begin{abstract}
Source code documentation plays an important role in software quality activities, such as code review, program comprehension, maintenance, and bug fixing. However, it is often fragmented across individual code elements and focuses on local implementation details, providing little information about their surrounding context. This limits both developers' understanding of software systems and the effectiveness of Large Language Models (LLMs), whose reasoning is constrained by finite context windows. Enriching documentation with hierarchical context has the potential to improve software comprehension and support downstream, quality-related tasks for both developers and LLM-based tools.
In this paper, we present an \emph{in silico} study that showcases the impact of augmenting context on the generation of code-level documentation, using three LLMs (GPT-4o-mini, Gemini-3-Flash, and DeepSeek v3.2). By varying the function, class, and module hierarchy visible to the LLMs, we assess whether this variation produces a significant difference in the resulting code documentation compared with the existing documentation, which serves as a proxy for developer intent. Our results show that Documentation-Aware prompts with class or module-level context consistently yield summaries that are closest to the original documentation. Adding surrounding documentation yields statistically significant improvements over the baseline across models, with effect sizes ranging from mostly weak (DeepSeek and GPT-4o-mini) to medium (Gemini). We restrict the study to this immediate hierarchical context; call-graph traversal and cross-module dependency context fall outside our scope and are left to future work. These findings provide a clear path to improve automated documentation tools and enhance LLMs' ability to reason about codebases for downstream tasks, such as code retrieval.
\end{abstract}


\renewcommand{\shortauthors}{Petrachini et al.}
\renewcommand{\shorttitle}{Context Matters: Improving LLM-Based Code Summarization}

\frenchspacing
\maketitle
\section{Introduction}
During software evolution, developers spend over half their working time on program comprehension \cite{XIA2094}, yet few software projects adequately comment their code \cite{10.1145/1858996.1859006}. In this context, source code documentation plays an important role in quality-related activities, including program comprehension, code review, and bug fixing. Furthermore, it has become of utmost relevance with the widespread adoption of generative artificial intelligence in the development process, where automated code generation needs to be verified and validated frequently, and likely reduces code ownership \cite{martinlopez2026morecode}.

When such documentation exists, it is often mismatched with the source code, missing, or outdated \cite{sym14030471}. Code summarization aims to close this gap: given a method's signature and body, Sridhara et al.\ \cite{10.1145/1858996.1859006} define the task as generating natural language text that "summarizes the overall actions of [the method] accurately, adequately, and concisely," producing descriptive comments that convey a method's intent through its major algorithmic actions to aid program comprehension and software maintenance.

The field has progressed from information retrieval and heuristic methods to abstract syntax tree embeddings \cite{alon2018code2veclearningdistributedrepresentations}, specialized transformer architectures such as CodeBERT \cite{feng2020codebertpretrainedmodelprogramming}, and hierarchical representations \cite{ZHOU2022106761}. Large Language Models (LLMs), trained on vast amounts of code and natural language data and viewable as learning a mapping function $f: C \mapsto D$ from source code $C$ to a natural language description $D$, have surpassed these approaches on standard benchmarks \cite{qin2023chatgptgeneralpurposenaturallanguage, sun2024sourcecodesummarizationera}, and developers now use LLM-based summarization in their workflows through products such as GitHub Copilot. Choosing how much surrounding context to provide is therefore a concrete design decision for anyone building documentation tools.

Current evaluations of LLM-based code summarization focus on whether models produce high-scoring summaries according to metrics such as BLEU \cite{Papineni2002BLEU}, METEOR \cite{Banerjee2005METEOR}, ROUGE \cite{Lin2004ROUGE}, and human or LLM-based judgment. They do not measure how the presence of existing documentation and varying levels of structural context influence whether the generated output approximates developer-written references, which remain the standard ground truth in code summarization evaluation \cite{sym14030471}.

Ahmed et al.\ \cite{ahmed2024automaticsemanticaugmentation} augment prompts with semantic information at the function level but do not investigate broader hierarchical context. Makharev and Ivanov \cite{11028203} explore class and repository-level context but rely on retrieval-augmented generation and few-shot learning, which conflates the effects of intrinsic code structure with externally retrieved information, preventing attribution of gains to either source. Without controlled evidence on how hierarchical scope affects summary quality, tool designers face an uninformed trade-off: including too much context wastes inference budget, while including too little risks producing summaries that miss the neighboring methods, shared attributes, and module-level definitions the target function depends on. There is no empirical basis to guide this decision.

By measuring how closely generated summaries match devel\-oper-written references at each context level, we identify which layers of context improve quality and which add cost without benefit. We investigate whether expanding hierarchical context from function-only to class or module level improves similarity to developer-written references (RQ1), whether preserving surrounding documentation adds value beyond code structure alone (RQ2), and whether these effects vary across three models, GPT-4o-mini, Gemini-3-Flash, and DeepSeek v3.2 (RQ3); the Research Method section states these questions in full.

We expand prior work by exploring a multi-layered context that includes the function itself, the class, and the module\footnote{Within the scope of this paper, which experiments on Python source codes, a module refers to a single source file (in this case, files with the .py extension)} to which it belongs, to produce hierarchically informed code documentation. We focus on the immediate hierarchical context (class and module definitions), which exposes the neighboring methods, shared attributes, and module-level definitions the target uses, rather than inter-procedural call graphs, cross-module data flow, or external retrieval-augmented generation (RAG) methods. By isolating the hierarchical scope, we establish a baseline for how well LLMs utilize the intrinsic structure of the codebase before introducing the complexity of dependency-aware context and external retrieval systems, which we leave to future work.

This distinction matters for tool designers: if hierarchical code structure alone drives quality improvements, tools can simplify the input pipeline by omitting surrounding documentation. The remainder of this paper is organized as follows: Section 2 positions the study in relation to related work, Section 3 describes the experimental method, and Section 4 presents the results.

\section{Background And Related Work}

Recent advances in code summarization have explored context augmentation and prompting strategies. We organize prior work along these two dimensions, paying attention to how each study treats pre-existing documentation in the surrounding scope.

\subsection{Context Augmentation For Code Summarization}

The most directly related line of work investigates how additional context beyond the bare function body improves generated summaries. Ahmed et al.~\cite{ahmed2024automaticsemanticaugmentation} augmented LLM prompts with syntactic elements extracted via program analysis (tagged identifiers, data-flow graphs, and repository metadata) and reported significant metric improvements over baselines. Their augmentation targets lexical and structural representations at the function level, but they do not investigate the effect of supplying the enclosing class or module definitions, nor do they examine whether pre-existing documentation in the surrounding scope influences the generated summary.

Lomshakov et al.~\cite{lomshakov2024proconsul} introduced \textsc{ProConSuL}, a framework that constructs project-level context for C/C++ code summarization by building a call graph and recursively summarizing callees. Their pipeline combines supervised fine-tuning with ORPO preference alignment and evaluates on 25 test functions. \textsc{ProConSuL} gathers context from other files via call-graph traversal, a program-analysis-based approach distinct from similarity-based retrieval. Their work does not examine the effect of intra-file hierarchical context or vary the presence of surrounding documentation.

Makharev and Ivanov~\cite{11028203} examined code summarization beyond the function level, evaluating LLMs with class-level and repository-level context on Python benchmarks (1,200 functions). At the class level, they provide the enclosing class body directly to the model; at the repository level, they supply RAG-retrieved code chunks. Their results show that few-shot learning and RAG-retrieved context enhance LLM performance. At the class level, the input representation overlaps: both approaches supply the enclosing class code.

Two gaps remain. Makharev and Ivanov do not distinguish between settings in which the surrounding documentation (docstrings of neighboring methods, the class-level docstring) is present versus stripped. They also do not extend the hierarchy to module-level scope (the full file), and their function-level and repository-level experiments rely on few-shot prompting and RAG, respectively.

\subsection{Prompting Strategies, Evaluation, And The Role Of Function Names}

A second line of work asks how prompting techniques and evaluation methods, rather than input content, affect summary quality. Sun et al.~\cite{sun2024sourcecodesummarizationera} conducted a systematic study of LLM-based code summarization, comparing five prompting techniques (zero-shot, few-shot, chain-of-thought, critique, and expert). They found that advanced prompting strategies do not consistently outperform simple zero-shot prompting, and that GPT-4-based evaluation aligns most closely with human judgment. Their study varies the prompting strategy while holding the input context constant at the function level. If zero-shot prompting already captures most gains from prompt engineering, the remaining lever for practitioners is the scope and composition of the input context itself.

Crupi et al.~\cite{crupi2025effectiveness} investigated the reliability of the LLM-as-a-Judge approach for code generation and summarization. When comparing judgments from multiple LLMs against human assessments, they found that GPT-4-turbo achieves moderate-to-substantial agreement with humans on summary content adequacy, whereas smaller models perform poorly as judges.

Haldar and Hockenmaier~\cite{haldar-hockenmaier-2024-analyzing} found that LLMs rely heavily on function names rather than on code structure when generating summaries, and that metrics such as BLEU and BERTScore provide largely redundant assessments. If models lean on function names as a shortcut, richer surrounding context may reduce that reliance, but no existing study tests this hypothesis directly.

\subsection{Positioning Of This Work}

The studies above leave unanswered whether improvements from the broader context are driven by additional code structure or by documentation already present in that structure. All prior work supplies context with documentation intact and does not test what happens when it is stripped. Makharev and Ivanov~\cite{11028203} test class-level context but do not vary the presence of documentation, and do not extend to the module scope. Our study addresses this gap by crossing context granularity (Function, Class, Module) with documentation mode, Documentation-Aware (DA) vs.\ Documentation-Independent (DI), as independent factors across multiple LLMs and 6,932 methods from 12 mature Python repositories.

\section{Research Method}
The goal of this study is to investigate how the presence of existing documentation and varying levels of context affect LLMs' ability to generate documentation that approximates developer intent.

\subsection{Research Questions}
\begin{itemize}
    \item \textbf{RQ1:} Does expanding hierarchical context from function-only to class or module level improve the similarity of LLM-generated documentation to developer-written references?
    \item \textbf{RQ2:} Does preserving surrounding documentation add value beyond code structure alone?
    \item \textbf{RQ3:} Do these effects vary across models?
\end{itemize}

\subsection{Study Design}
While human-written references often lack the stylistic completeness of LLM outputs, they remain the source of ground truth for semantic alignment. Therefore, while also including $n$-gram metric analysis, we use BERTScore to measure semantic alignment, and LLM-as-a-Judge to evaluate informational completeness.
Additionally, this study examines whether retaining the original reference documentation affects the metrics and keywords used in the generated summaries.

Following guidelines for empirical studies utilizing LLMs~{\cite{baltes2025guidelinesempiricalstudiessoftware}} and empirical standards for data science experiments \cite{ralph2020empirical}, we use an incomplete factorial design with five experimental conditions, controlled by two independent factors. The first factor, \textbf{context granularity}, controls the hierarchical scope of code visible to the model (Function, Class, or Module). The second factor, \textbf{documentation mode}, controls whether existing documentation in the surrounding scope is retained or removed. With this approach, context granularity determines how much code structure the model sees, while documentation mode determines whether the docstrings within that structure are preserved.

We define \textit{Documentation-Aware} (DA) as an input where the target function's docstring is removed, but documentation in the surrounding scope (class or module) is preserved. \textit{Documentation-Independent} (DI) involves stripping all documentation from the input entirely.

At the function-only level, no surrounding scope is provided to the model. Therefore, the distinction between Documentation-Aware (DA) and Documentation-Independent (DI) is not applicable. We treat the "Function-only" condition as the baseline representing standard code summarization without hierarchical context. This yields five experimental scenarios: the Function-only baseline, plus Class and Module contexts under both DI and DA conditions, illustrated in Figure~\ref{fig:running_example}.

These conditions map directly to our research questions. RQ1 is addressed by comparing the Function-only baseline against Class and Module conditions, isolating the effect of hierarchical scope. RQ2 is addressed by comparing DI against DA at each context level, isolating the effect of surrounding documentation. RQ3 is addressed by conducting these comparisons separately for each of the three LLMs.

To illustrate the five conditions, Figure~\ref{fig:running_example} shows an abbreviated module from the \texttt{requests} library. Colors indicate which scope includes each region; underlined docstrings are preserved in DA conditions but stripped in DI.

\definecolor{modscope}{RGB}{46,139,87}
\definecolor{clsscope}{RGB}{30,90,180}
\definecolor{tgtscope}{RGB}{140,20,20}

\begin{figure}[htbp]
\begin{lstlisting}[
  language=Python,
  basicstyle=\small\ttfamily,
  frame=single,
  framerule=0.4pt,
  escapechar=|
]
|\textcolor{modscope}{\underline{"""requests.structures}}|
|\textcolor{modscope}{\underline{Data structures..."""}}|
|\textcolor{modscope}{import os, collections}|
|\textcolor{modscope}{from itertools import islice}|

|\textcolor{modscope}{class IteratorProxy(object): ...}|

|\textcolor{clsscope}{class CaseInsensitiveDict(}|
|\textcolor{clsscope}{~~~~~~~~collections.MutableMapping):}|
    |\textcolor{clsscope}{\underline{"""A case-insensitive dict-like}}|
    |\textcolor{clsscope}{\underline{object..."""}}|
    |\textcolor{clsscope}{def \_\_init\_\_(self, data=None, **kw):}|
    |\textcolor{clsscope}{~~~~self.\_store = dict()}|
    |\textcolor{clsscope}{~~~~...}|
    |\textcolor{clsscope}{def \_\_setitem\_\_(self, key, value):}|
    |\textcolor{clsscope}{~~~~self.\_store[key.lower()] = (key, value)}|
    |\textcolor{clsscope}{...}|
    |\textcolor{tgtscope}{\textbf{def lower\_items(self):}}|
    |\textcolor{tgtscope}{~~~~\textbf{<< docstring removed >>}}|
    |\textcolor{tgtscope}{\textbf{~~~~return ((lowerkey, keyval[1])}}|
    |\textcolor{tgtscope}{\textbf{~~~~~for lowerkey, keyval}}|
    |\textcolor{tgtscope}{\textbf{~~~~~in self.\_store.items())}}|
    |\textcolor{clsscope}{...}|

|\textcolor{modscope}{class LookupDict(dict): ...}|
\end{lstlisting}
\vspace{2pt}
{\small
\textcolor{tgtscope}{$\blacksquare$} Target function (all conditions)\quad
\textcolor{clsscope}{$\blacksquare$} Class scope (Class + Module)\quad
\textcolor{modscope}{$\blacksquare$} Module scope (Module only)\quad
\underline{Underlined} = docstring (DA only, stripped in DI)}
\caption{The five experimental conditions applied to \texttt{lower\_items} from \texttt{requests/structures.py}. Ellipses denote truncated code. The target function's docstring is removed in all conditions.}
\label{fig:running_example}
\Description{Color-coded Python module listing showing three scope levels: module (green), class (blue), and target function (red), with underlined docstrings indicating DA-only content.}
\end{figure}

\subsection{Dataset}
We sourced 12 Python open-source repositories from the SWE-Bench dataset~\cite{jimenez2024swebenchlanguagemodelsresolve}, chosen for their maturity and active developer communities. Only artifacts documented at all three hierarchical levels (function, class, and module) were retained, ensuring consistency in metrics derived from context augmentation across experimental conditions. Table~\ref{tab:repos-commits} lists the repositories and their pinned commits, along with descriptive statistics per repository.

\begin{table*}[!t]
\centering
\small
\caption{Studied repositories: pinned commits and per-repository descriptive statistics. Modules, Classes, and Functions columns count only those with docstrings. "After Filter" is the count of functions documented at all three hierarchical levels.}
\label{tab:repos-commits}
\footnotesize
\begin{tabular}{llrrrrr}
\toprule
\textbf{Repository} & \textbf{SHA Commit} & \textbf{Modules} & \textbf{Classes} & \textbf{Functions} & \textbf{Docstrings} & \textbf{After Filter} \\
\midrule
astropy      & \texttt{d16bfe05} & 227 & 978 & 2,328 & 3,533 & 1,205 \\
django       & \texttt{e7fd69d0} & 153 & 606 & 2,153 & 2,912 & 480 \\
flask        & \texttt{d8c37f43} & 1 & 42 & 154 & 197 & 10 \\
matplotlib   & \texttt{b7d05919} & 104 & 543 & 2,398 & 3,045 & 1,798 \\
pylint       & \texttt{182cc539} & 66 & 162 & 577 & 805 & 343 \\
pytest       & \texttt{6995257c} & 13 & 19 & 286 & 318 & 1 \\
requests     & \texttt{110048f9} & 20 & 65 & 164 & 249 & 77 \\
scikit-learn & \texttt{b90661d6} & 169 & 385 & 864 & 1,418 & 750 \\
seaborn      & \texttt{94621cef} & 16 & 69 & 263 & 348 & 78 \\
sphinx       & \texttt{7bdc11e8} & 169 & 395 & 465 & 1,029 & 380 \\
sympy        & \texttt{8dcb12a6} & 306 & 938 & 3,018 & 4,262 & 1,794 \\
xarray       & \texttt{863e4906} & 22 & 92 & 405 & 519 & 16 \\
\midrule
\textbf{Total} & & \textbf{1,266} & \textbf{4,294} & \textbf{13,075} & \textbf{18,635} & \textbf{6,932} \\
\bottomrule
\end{tabular}
\end{table*}

\begin{figure}[!t]
  \centering
  \includegraphics[height=0.8\textheight,keepaspectratio]{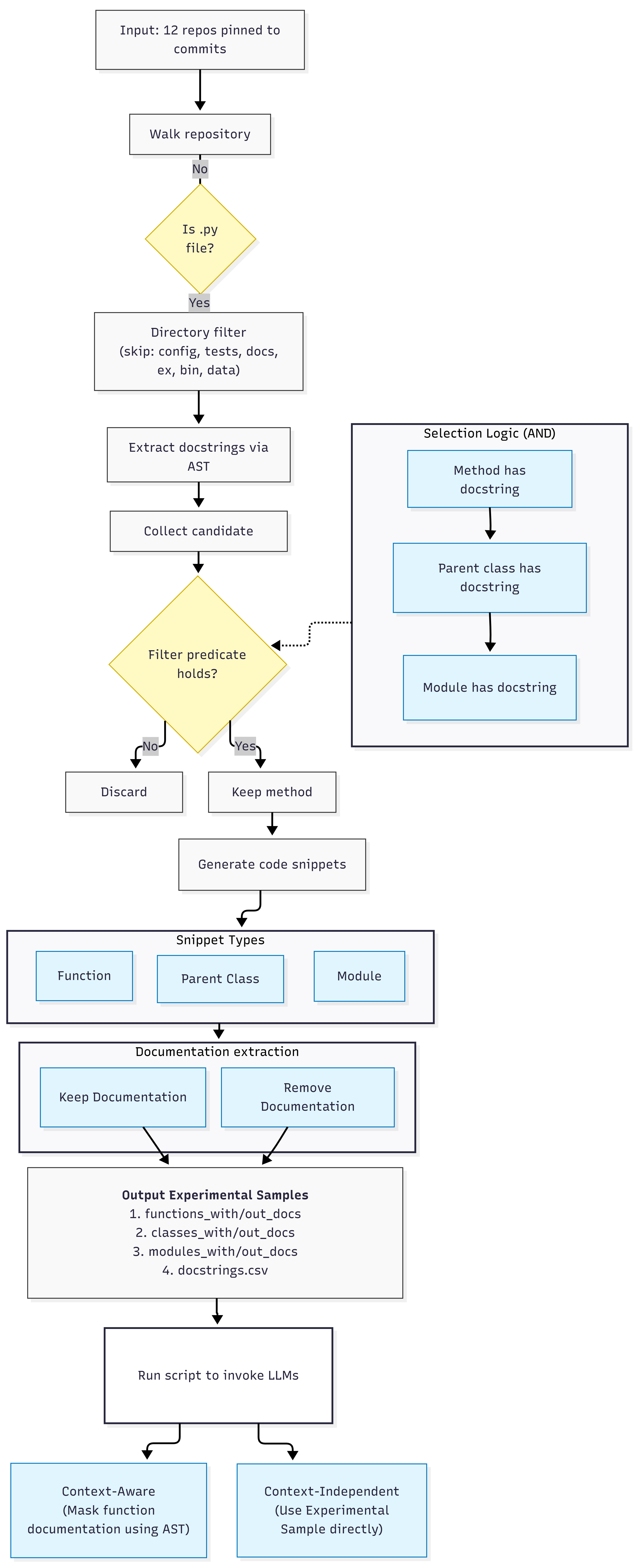}
  \caption{Experimental sample pipeline procedure}
  \label{fig:pipeline_procedure}
  \Description{This figure shows the data pipeline of the experiment.}
\end{figure}

To ensure a consistent basis for comparison across experimental conditions, we filtered the dataset, excluding modules that were not originally documented at all three hierarchical levels. To simplify the extraction process, we excluded nested class definitions. After filtering, the dataset consisted of 6,932 methods across the 12 repositories. We excluded 4 samples from metric evaluation because their reference docstrings were empty or consisted solely of numeric values, yielding 6,928 samples for the evaluation reported in Section~\ref{statistical_analysis}.
Across the dataset, modules contain a median of two classes (Q1=1, Q3=4), indicating that class-level and module-level contexts differ in scope.
Figure~\ref{fig:pipeline_procedure} shows the architecture of the data preparation pipeline.
The median module size is 517 LoC (IQR: 239--993), and the median reference docstring length is 94 characters (IQR: 47--273).

\subsection{Experimental Procedure Setup}
To generate code summaries, we used the models listed in Table \ref{tab:llm_models}, setting the temperature to 0 and top\_p to 1.
\begin{table}[!t]
      \caption{Large Language Models used in the experiment}
  \label{tab:llm_models}
  \centering
  \begin{tabular}{llc}
    \toprule
    \textbf{Provider} & \textbf{Model ID} & \textbf{Knowledge Cutoff} \\
    \midrule
    OpenAI   & gpt-4o-mini-2024-07-18 & Oct 2023 \\
    Google   & gemini-3-flash-preview & Jan 2025 \\
    DeepSeek  & Deepseek v3.2         & Mar 2025 \\
    \bottomrule
  \end{tabular}
\end{table}
We employed low-cost and general-purpose API models from three different providers, including DeepSeek v3.2, which has open weights. The generation prompt does not enforce a specific docstring style, allowing each model to produce documentation in whatever format it deems appropriate for the function. The full generation prompt is in the replication package.

Two input conditions were prepared for each artifact: (1) with existing contextual documentation retained, but removing the target documentation, and (2) with all documentation removed. In both cases, the same source code context was supplied to the LLM, ensuring comparability across conditions. The experiment ran in a WSL2 environment on Windows using Python 3.11. Generation invoked the OpenAI, Gemini, and DeepSeek APIs, whereas the LLM-as-a-Judge evaluation used the Claude API from a model family different from those of the generators to avoid self-preference bias (Section~\ref{evaluation}). BERTScore was computed using an NVIDIA RTX 4090 24GB with the google/mt5-small model as the underlying encoder (this model is used solely for the BERTScore computation). No truncation was required, as the longest (prompt + context) input was 48,568 tokens, as counted by tiktoken using the cl100k\_base encoding.

\subsection{Evaluation}
\label{evaluation}

\textbf{N-gram based metrics.} An $n$-gram is a contiguous sequence of tokens that captures ordering and co-occurrence between consecutive words. BLEU \cite{Papineni2002BLEU} computes $n$-gram precision between candidate and reference with a brevity penalty, ROUGE \cite{Lin2004ROUGE} measures the ROUGE-1 F1 score of unigram overlap between candidate and reference, and METEOR \cite{Banerjee2005METEOR} combines unigram precision and recall with stemming and synonym matching.

\textbf{Similarity-based metrics.} While $n$-gram metrics capture syntactic overlap, they often fail to recognize semantic equivalence when the same idea is expressed with different wording. BERTScore \cite{zhang2020bertscore} addresses this by encoding candidate and reference texts into dense vector representations and computing cosine similarity in the embedding space. We compute BERTScore using the google/mt5-small encoder. We selected this encoder for two reasons: it shows a reasonable correlation with human judgments (0.64 on the WMT16 To-English task, as reported in the BERTScore model-selection sheet), and it supports a large input token limit, which helps accommodate longer docstrings without truncation.

\textbf{LLM-as-a-Judge.} Zheng et al. \cite{10.5555/3666122.3668142} proposed the LLM-as-a-Judge framework to address the scalability constraints of human annotation and the semantic limitations of $n$-gram metrics. Among the evaluation protocols they establish, pairwise comparison has been shown to yield more stable and distinguishable results by grounding the evaluation in relative preference rather than absolute scoring.

Once the summaries are generated, we compute BLEU, ROUGE, METEOR, and BERTScore using the Evaluate library \cite{von-werra-etal-2022-evaluate} and compare them against the original reference documentation. Before computing the metrics, we extract the docstring from each generated output and remove Markdown code fences, triple quotes, indentation, and surrounding whitespace from the generated and reference docstrings; each metric then uses its default tokenization. Statistical testing follows the procedure described in Section~\ref{sec:stat_plan}.

To mitigate the limitations of $n$-gram metrics and their potential bias toward verbose summaries, we complemented our evaluation with an LLM-as-a-Judge assessment. In this setup, a separate LLM acts as an automated evaluator: it receives a function's source code, its surrounding context, and three candidate comments, then produces a strict ranking from best to worst. The three candidates are the reference docstring extracted from the repository, the DI-generated comment, and the DA-generated comment.

For each evaluation instance, the mapping between candidates and the labels A, B, C is determined by a random permutation, drawn once per function and context level to mitigate position bias~\cite{10.5555/3666122.3668142}. The judge is prompted to return a JSON object containing a ranked list such as \texttt{["B","A","C"]} and a short rationale. We extract the ranking from the response and convert it to numerical ranks (1=best, 3=worst) for each candidate. The prompt enforces strict ordering with no ties. It instructs the judge to treat the function body as primary evidence and the surrounding context only as secondary support for disambiguating intent, penalizing claims not supported by the code (the full judge prompt is in the replication package).

We adopted three evaluation criteria from prior work on LLM-based code summarization assessment: content adequacy, conciseness, and fluency and understandability. Crupi et al.~\cite{crupi2025effectiveness} validated these criteria against human judgments for code summarization and code generation tasks, finding that GPT-4-based rankings under these criteria achieved high agreement with expert annotators. Sun et al.~\cite{sun2024sourcecodesummarizationera} independently confirmed that GPT-4 evaluation aligned most closely with human assessment among several automated metrics they tested. Adequacy measures whether the comment captures what the function does, conciseness penalizes unnecessary detail, and fluency assesses whether a developer can read the comment without difficulty. Together, they cover complementary dimensions of comment quality.

Self-preference bias refers to the tendency of an LLM judge to systematically favor outputs produced by its own model family over outputs of comparable quality from other models~\cite{chen-etal-2025-beyond}. To mitigate this risk, we use \texttt{claude-sonnet-4-6} as the judge, a model from a different provider and family than all three generators in our study. The judge was queried with a temperature of 0 to maximize reproducibility. We sampled 1000 instances for evaluation, stratified by reference docstring length into quartiles (250 instances per quartile) using a fixed random seed.

\textbf{Human calibration of the judge.} We calibrated the judge on a stratified sample of 20 tasks covering all six model and context-level instances, ranked blind by one author under the same criteria as the judge, with a second blinded author as an inter-rater baseline. The judge agreed with each author on the DA versus DI comparison underlying RQ2 in 65--70\% of tasks (Cohen's kappa 0.30--0.40) and matched the raters' consensus choice in 11 of the 15 tasks on which they aligned (73\%). Over the full rankings (60 rank assignments), the judge agreed with each rater (Krippendorff ordinal alpha 0.33 and 0.41) more than the raters agreed with each other (alpha 0.11), so the judge ranks candidate summaries at least as consistently as the raters themselves. This supports relying on the judge for the aggregate DA versus DI comparisons. The rating sheets are in the replication package.

\subsection{Statistical Tests}
\label{sec:stat_plan}
We first test for normality using the D'Agostino-Pearson test. If normality is rejected, we apply a non-parametric Kruskal-Wallis test to assess whether each metric exhibits the same distribution across context levels and documentation scenarios, at a significance level of 0.05. Post-hoc pairwise comparisons use the Dunn-Bonferroni test, and we report effect sizes using Pearson's $r = |Z|/\sqrt{N}$, benchmarked with Cohen's guidelines (r = .10 weak, .30 medium, .50 strong) following Kampenes et al.~\cite{kampenes2007systematic}. The statistics were calculated using IBM SPSS Statistics software~\cite{spss_software_2025}.

\section{Results}
\subsection{Overview}

Table~\ref{tab:desc_length_stats} summarizes the generated description lengths across all five experimental conditions. The median values range from 586 to 674 characters. Descriptions produced with broader context tend to be longer, with Module DA producing the highest median (674 characters).

\begin{table}[!t]
\centering
\small
\caption{Generated description length (characters) across experimental conditions, pooled over all models.}
\label{tab:desc_length_stats}
\footnotesize
\begin{tabular}{lrrrr}
\toprule
\textbf{Condition} & \textbf{Median} & \textbf{Q1} & \textbf{Q3} & \textbf{IQR} \\
\midrule
Function-only (Baseline) & 586 & 334 & 960 & 626 \\
Class DI                 & 641 & 365 & 1009 & 644 \\
Class DA                 & 622 & 334 & 988 & 654 \\
Module DI                & 662 & 367 & 1042 & 675 \\
Module DA                & 674 & 362 & 1046 & 684 \\
\bottomrule
\end{tabular}
\end{table}

The median generated docstring length varies by model: Gemini produces the shortest outputs (median 312 characters at baseline), GPT-4o-mini produces mid-length outputs (median 586), and DeepSeek produces the longest (median 1029). The pooled median (586 characters) is roughly six times the median reference docstring (94 characters).

We report the metrics from comparing generated docstrings against the original references. Table \ref{tab:overall_mean_std} summarizes the pooled statistics (mean $\pm$ std) for each metric across all scenarios. Per-model results are available in the replication package; the pooled table highlights the overall effect across models.

\begin{table*}[!t]
\centering
\small
\caption{Overall pooled performance across all generated instances (mean $\pm$ std) by context level and condition.}
\label{tab:overall_mean_std}
\begin{tabular}{@{}lccccc@{}}
\toprule
\textbf{Metric} &
\textbf{Function-only (Baseline)} &
\textbf{Class DI} &
\textbf{Class DA} &
\textbf{Module DI} &
\textbf{Module DA} \\
\midrule
BERTScore & 0.5566 $\pm$ 0.1157 & 0.5683 $\pm$ 0.1288 & \textbf{0.6182} $\pm$ 0.1669 & 0.5745 $\pm$ 0.1353 & 0.6124 $\pm$ 0.1651 \\
BLEU      & 0.0611 $\pm$ 0.1393 & 0.0810 $\pm$ 0.1681 & \textbf{0.1738} $\pm$ 0.2771 & 0.0926 $\pm$ 0.1848 & 0.1664 $\pm$ 0.2696 \\
ROUGE     & 0.2482 $\pm$ 0.1897 & 0.2649 $\pm$ 0.2111 & \textbf{0.3400} $\pm$ 0.2758 & 0.2731 $\pm$ 0.2211 & 0.3286 $\pm$ 0.2714 \\
METEOR    & 0.2831 $\pm$ 0.1814 & 0.3058 $\pm$ 0.1993 & \textbf{0.4028} $\pm$ 0.2631 & 0.3176 $\pm$ 0.2098 & 0.3955 $\pm$ 0.2592 \\
\bottomrule
\end{tabular}
\end{table*}

Across pooled results, adding hierarchical context improves scores relative to the function-only baseline. Documentation-Aware (DA) consistently outperforms Documentation-Independent (DI) at both class and module levels. The largest gains are observed for DA with class context, which achieves the highest pooled values across all four metrics. Class DA and Module DA produce similar scores for Gemini and GPT-4o-mini, where differences do not reach statistical significance. For DeepSeek, Class DA significantly exceeds Module DA on BERTScore, ROUGE, and METEOR.

\subsection{Statistical Analysis}
\label{statistical_analysis}
Normality is rejected for all metric distributions across all models and scenarios (D'Agostino-Pearson, $p \le 5.44\times10^{-6}$), so we apply the non-parametric tests described in Section~\ref{sec:stat_plan}.
\begin{table}[!t]
\centering
\small
\setlength{\tabcolsep}{4pt}
\caption{Kruskal--Wallis tests across experimental groups, reported by model. Effect size computed as $\eta_H^2=(H-k+1)/(n-k)$ with $k=5$ groups and $n=34640$ per model.}
\label{tab:kruskal_wallis_by_model}
\footnotesize
\renewcommand{\arraystretch}{0.92}
\begin{tabular}{llrrr}
\toprule
Model & Metric & $H$ & $\eta^{2}_{H}$ & $p$ \\
\midrule
deepseek v3.2 & BERTScore & 226.354 & 0.006 & $<$.001 \\
deepseek v3.2 & BLEU      & 677.710 & 0.019 & $<$.001 \\
deepseek v3.2 & ROUGE     & 215.258 & 0.006 & $<$.001 \\
deepseek v3.2 & METEOR    & 443.108 & 0.013 & $<$.001 \\
\midrule
gemini-3-flash-preview & BERTScore & 2501.200 & 0.072 & $<$.001 \\
gemini-3-flash-preview & BLEU      & 2587.016 & 0.075 & $<$.001 \\
gemini-3-flash-preview & ROUGE     & 2174.908 & 0.063 & $<$.001 \\
gemini-3-flash-preview & METEOR    & 2785.378 & 0.080 & $<$.001 \\
\midrule
gpt-4o-mini & BERTScore & 428.818 & 0.012 & $<$.001 \\
gpt-4o-mini & BLEU      & 1597.195 & 0.046 & $<$.001 \\
gpt-4o-mini & ROUGE     & 197.029 & 0.006 & $<$.001 \\
gpt-4o-mini & METEOR    & 1371.514 & 0.039 & $<$.001 \\
\bottomrule
\end{tabular}
\end{table}
Kruskal--Wallis tests showed significant differences across the five experimental conditions for every metric, as in Table \ref{tab:kruskal_wallis_by_model}. We follow up with a post-hoc Dunn-Bonferroni test to identify which of the groups had significant deviations, by computing the effect size statistic r and then benchmark it using Cohen's guidelines of Pearson's r = .10 (weak effect), .30 (medium effect), and .50 (strong effect), in line with existing literature in software engineering \cite{kampenes2007systematic}.

Table \ref{tab:DI_effect_table} shows the effect sizes for DI metrics, per level of context and model. For DeepSeek and GPT-4o-mini, all effect sizes are negligible (r < .10). Gemini stands out: at the module level, Baseline vs. DI reaches weak effects (r = 0.13--0.15), indicating that code structure alone produces measurable gains for this model. At the class level, Gemini DI effects remain negligible but approach the weak threshold (r = 0.09--0.10).

\begin{table}[!t]
\centering
\caption{Function-only (Baseline) vs. Documentation-Independent (DI) effect sizes ($r=|Z|/\sqrt{N}$, with $N=13856$). Most contrasts are statistically significant (Dunn--Bonferroni adjusted p < .001). All effect sizes are negligible (r < .10) for DeepSeek and GPT-4o-mini, Gemini Module DI reaches the weak threshold. Effect labels: Neg. $<.10$, Weak $.10$--$.30$, Mod. $\ge .30$.}
\label{tab:DI_effect_table}
\footnotesize
\setlength{\tabcolsep}{3.5pt}
\renewcommand{\arraystretch}{0.9}
\begin{tabular}{l l r l}
\toprule
Comparison & Metric & r & Effect \\
\midrule
\multicolumn{4}{l}{\textbf{DeepSeek v3.2}}\\
\textbf{Baseline vs.\ DI Class}  & BERTScore & 0.005 & Neg. \\
                                & BLEU      & 0.041 & Neg. \\
                                & ROUGE     & 0.012 & Neg. \\
                                & METEOR    & 0.018 & Neg. \\
\addlinespace
\textbf{Baseline vs.\ DI Module} & BERTScore & 0.008 & Neg. \\
                                & BLEU      & 0.058 & Neg. \\
                                & ROUGE     & 0.019 & Neg. \\
                                & METEOR    & 0.018 & Neg. \\
\midrule

\multicolumn{4}{l}{\textbf{Gemini-3-Flash}}\\
\textbf{Baseline vs.\ DI Class}  & BERTScore & 0.097 & Neg. \\
                                & BLEU      & 0.098 & Neg. \\
                                & ROUGE     & 0.090 & Neg. \\
                                & METEOR    & 0.099 & Neg. \\
\addlinespace
\textbf{Baseline vs.\ DI Module} & BERTScore & 0.145 & Weak \\
                                & BLEU      & 0.146 & Weak \\
                                & ROUGE     & 0.132 & Weak \\
                                & METEOR    & 0.150 & Weak \\
\midrule

\multicolumn{4}{l}{\textbf{GPT-4o-Mini}}\\
\textbf{Baseline vs.\ DI Class}  & BERTScore & 0.008 & Neg. \\
                                & BLEU      & 0.050 & Neg. \\
                                & ROUGE     & 0.001 & Neg. \\
                                & METEOR    & 0.032 & Neg. \\
\addlinespace
\textbf{Baseline vs.\ DI Module} & BERTScore & 0.013 & Neg. \\
                                & BLEU      & 0.068 & Neg. \\
                                & ROUGE     & 0.001 & Neg. \\
                                & METEOR    & 0.046 & Neg. \\
\bottomrule
\end{tabular}
\end{table}

Table \ref{tab:DA_effect_table} shows the effect sizes for DA per level of context. All Baseline vs. DA contrasts are statistically significant (p < .001). The magnitude varies by model. Gemini produces medium effect sizes across all four metrics (r = 0.31--0.36), the largest observed in this study. GPT-4o-mini shows mostly weak effects (r = 0.07--0.26), strongest on BLEU and METEOR. DeepSeek effects are mostly weak (r = 0.04--0.18), with only BLEU and METEOR reaching the weak threshold at both context levels.

\begin{table}[!t]
\centering
\caption{Function-only (Baseline) vs. Documentation-Aware (DA) effect sizes ($r=|Z|/\sqrt{N}$, with $N=13856$). All contrasts are statistically significant (p < .001). Effect labels: Neg. $<.10$, Weak $.10$--$.30$, Mod. $\ge .30$.}
\label{tab:DA_effect_table}
\footnotesize
\setlength{\tabcolsep}{3.5pt}
\renewcommand{\arraystretch}{0.9}
\begin{tabular}{l l r l}
\toprule
Comparison & Metric & r & Effect \\
\midrule
\multicolumn{4}{l}{\textbf{DeepSeek v3.2}}\\
\textbf{Baseline vs.\ DA Class}  & BERTScore & 0.100 & Weak \\
                             & BLEU      & 0.178 & Weak \\
                             & ROUGE     & 0.086 & Neg. \\
                             & METEOR    & 0.142 & Weak \\
\addlinespace
\textbf{Baseline vs.\ DA Module} & BERTScore & 0.069 & Neg. \\
                             & BLEU      & 0.162 & Weak \\
                             & ROUGE     & 0.042 & Neg. \\
                             & METEOR    & 0.106 & Weak \\
\addlinespace
\midrule
\multicolumn{4}{l}{\textbf{Gemini-3-Flash}}\\
\textbf{Baseline vs.\ DA Class}  & BERTScore & 0.345 & Mod. \\
                             & BLEU      & 0.347 & Mod. \\
                             & ROUGE     & 0.322 & Mod. \\
                             & METEOR    & 0.356 & Mod. \\
\addlinespace
\textbf{Baseline vs.\ DA Module} & BERTScore & 0.330 & Mod. \\
                             & BLEU      & 0.339 & Mod. \\
                             & ROUGE     & 0.306 & Mod. \\
                             & METEOR    & 0.354 & Mod. \\
\addlinespace
\midrule
\multicolumn{4}{l}{\textbf{GPT-4o-Mini}}\\
\textbf{Baseline vs.\ DA Class}  & BERTScore & 0.126 & Weak \\
                             & BLEU      & 0.252 & Weak \\
                             & ROUGE     & 0.085 & Neg. \\
                             & METEOR    & 0.230 & Weak \\
\addlinespace
\textbf{Baseline vs.\ DA Module} & BERTScore & 0.114 & Weak \\
                             & BLEU      & 0.255 & Weak \\
                             & ROUGE     & 0.066 & Neg. \\
                             & METEOR    & 0.224 & Weak \\
\bottomrule
\end{tabular}
\end{table}

Table \ref{tab:DI_DA_effect_table} shows the effect sizes for comparisons between DI and DA at each level of context. All DI vs. DA contrasts are statistically significant (p < .001). Gemini shows the largest DI-to-DA gains, with class-level effects reaching r = 0.23--0.26 across all metrics. GPT-4o-mini shows weak effects on BLEU (r = 0.20) and METEOR (r = 0.20), with negligible effects on ROUGE. DeepSeek effects are negligible to weak (r = 0.06--0.14). These results confirm that the benefit of preserving documentation is model-dependent, with Gemini deriving the largest gains.

Returning to RQ1, expanding the hierarchical context under DA prompts produces statistically significant improvements across all four metrics and all three models. Under DI prompts, effect sizes remain negligible for DeepSeek and GPT-4o-mini (r < .10) but reach weak levels for Gemini at module scope (r = 0.13--0.15). Under DA prompts, the effect is model-dependent: Gemini produces medium effect sizes across all four metrics (r = 0.31--0.36), GPT-4o-mini produces mostly weak effects (r = 0.07--0.26), and DeepSeek produces mostly weak effects (r = 0.04--0.18). Broader context increases similarity to developer-written references, with the practical magnitude ranging from mostly weak (DeepSeek) to medium (Gemini).

For RQ2, Documentation-Aware outperforms Documentation-Independent prompts at matched context levels, but the magnitude is model-dependent. Gemini shows the largest DI-to-DA effect sizes (r = 0.18--0.26), GPT-4o-mini shows weak effects on BLEU and METEOR (r = 0.18--0.20), and DeepSeek shows negligible to weak effects (r = 0.06--0.14). The LLM-as-a-Judge evaluation confirms the direction for every model: DA is significantly preferred over DI at all model and context-level combinations (r = 0.08--0.16, p $\le$ .016), with the judge effect sizes mostly weak even where the metric effects are medium. Preserving surrounding documentation provides a consistent benefit on both similarity metrics and judge preference.

\begin{table}[!t]
\centering
\caption{Documentation-Independent vs.\ Documentation-Aware (DI vs.\ DA) effect sizes for matched context levels ($r=|Z|/\sqrt{N}$, with $N=13856$). All shown DI vs.\ DA contrasts are statistically significant (p < .001). Effect labels: Neg. $<.10$, Weak $.10$--$.30$, Mod. $\ge .30$.}
\label{tab:DI_DA_effect_table}
\footnotesize
\setlength{\tabcolsep}{3.5pt}
\renewcommand{\arraystretch}{0.9}
\begin{tabular}{l l r l}
\toprule
Comparison & Metric & r & Effect \\
\midrule
\multicolumn{4}{l}{\textbf{DeepSeek v3.2}}\\
\textbf{DI Class vs.\ DA Class}   & BERTScore & 0.095 & Neg. \\
                                 & BLEU      & 0.137 & Weak \\
                                 & ROUGE     & 0.098 & Neg. \\
                                 & METEOR    & 0.124 & Weak \\
\addlinespace
\textbf{DI Module vs.\ DA Module} & BERTScore & 0.061 & Neg. \\
                                 & BLEU      & 0.105 & Weak \\
                                 & ROUGE     & 0.061 & Neg. \\
                                 & METEOR    & 0.088 & Neg. \\
\addlinespace
\midrule
\multicolumn{4}{l}{\textbf{Gemini-3-Flash}}\\
\textbf{DI Class vs.\ DA Class}   & BERTScore & 0.248 & Weak \\
                                 & BLEU      & 0.249 & Weak \\
                                 & ROUGE     & 0.232 & Weak \\
                                 & METEOR    & 0.257 & Weak \\
\addlinespace
\textbf{DI Module vs.\ DA Module} & BERTScore & 0.185 & Weak \\
                                 & BLEU      & 0.193 & Weak \\
                                 & ROUGE     & 0.175 & Weak \\
                                 & METEOR    & 0.204 & Weak \\
\addlinespace
\midrule
\multicolumn{4}{l}{\textbf{GPT-4o-Mini}}\\
\textbf{DI Class vs.\ DA Class}   & BERTScore & 0.118 & Weak \\
                                 & BLEU      & 0.202 & Weak \\
                                 & ROUGE     & 0.086 & Neg. \\
                                 & METEOR    & 0.198 & Weak \\
\addlinespace
\textbf{DI Module vs.\ DA Module} & BERTScore & 0.101 & Weak \\
                                 & BLEU      & 0.187 & Weak \\
                                 & ROUGE     & 0.067 & Neg. \\
                                 & METEOR    & 0.178 & Weak \\
\bottomrule
\end{tabular}
\end{table}

\subsection{LLM-as-a-Judge Analysis}
\label{sec:judge_analysis}
Table~\ref{tab:judge_win_rates} reports win rates from the Claude Sonnet 4.6 judge. The developer-written reference achieves the highest win rate for GPT-4o-mini (51.5--53.4\%) and DeepSeek (42.4--47.1\%), while Gemini references win under 10\% of their comparisons. Between the two generated candidates, DA beats DI consistently, confirmed by Wilcoxon signed-rank tests for every model and context level. For GPT-4o-mini, DA beats DI in 56.8\% of class comparisons (p < .0001, r = 0.134) and 56.1\% of module comparisons (p < .001, r = 0.109). For DeepSeek, DA beats DI in 57.3\% of class (p < .0001, r = 0.159) and 53.6\% of module comparisons (p = .016, r = 0.077). For Gemini, DA beats DI in 52.0\% of class (p < .001, r = 0.113) and 53.6\% of module comparisons (p < .001, r = 0.109). Figure~\ref{fig:judge_outputs} shows one complete evaluation.

\begin{figure*}[!t]
    \centering
    \includegraphics[width=0.7\textwidth]{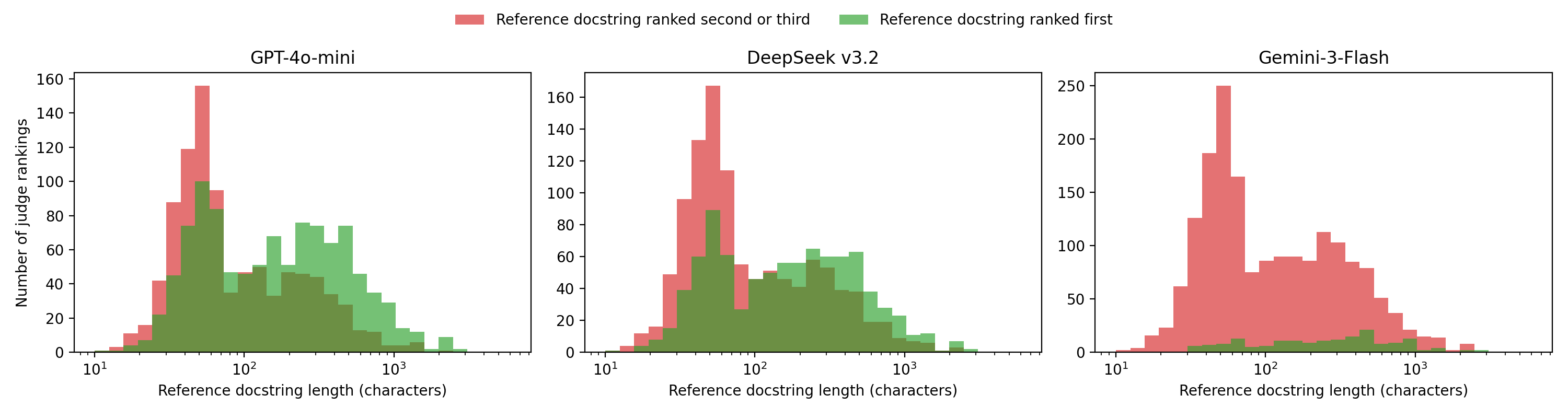}
    \caption{Reference vs LLM-generated comment wins by comment size}
    \label{fig:ref_winrate_by_length}
    \Description{This figure shows the distribution of reference wins over reference comment size, by model.}
\end{figure*}

We also find a statistically significant association between the length of the reference docstring and the reference win rate: as reference docstrings get longer, they are more likely to be ranked first. Figure~\ref{fig:ref_winrate_by_length} shows histograms of reference docstring lengths for cases where the reference is ranked first (green) versus ranked second/third (red). Table~\ref{tab:ref_rank1_mwu} reports the confirming Mann--Whitney U tests with length summary statistics. The judge does not simply prefer longer text. Generated candidates are about six times longer than references. However, the judge prefers references more as they get longer. This is the opposite of what would happen if the judge liked long text. The judge's reasons focus on information, not length. For example, the judge ranked a 526-character reference first because it ``correctly notes that the function does nothing'' when both label positions are equally free. Both generated candidates missed this edge case. Longer references win because they contain more of the information the judge verifies against the code.

\begin{figure}[!t]
\centering
\small
\begin{tabular}{@{}p{0.97\linewidth}@{}}
\toprule
\textbf{Function} (\texttt{requests}, \texttt{Response.apparent\_encoding}) \\
\texttt{@property} \\
\texttt{def apparent\_encoding(self):} \\
\texttt{\ \ return chardet.detect(self.content)['encoding']} \\
\midrule
\textbf{Reference} (ranked 2nd): "The apparent encoding, provided by the chardet library" \\
\addlinespace
\textbf{DA, module} (ranked 1st): "The apparent encoding, guessed by using the chardet library. The encoding is detected by analyzing the response content (bytes). This is used as a fallback when the 'encoding' attribute is None. Returns: str or None [...]" \\
\addlinespace
\textbf{DI, module} (ranked 3rd): "The apparent encoding of the response content, as detected by chardet. [...] It is particularly useful when the server does not specify an encoding in the Content-Type header [...] For large responses, consider alternative approaches to avoid loading the entire content into memory. [...]" \\
\addlinespace
\textbf{Judge rationale}: the DA summary "accurately captures the core behavior [...] the return type (str or None), and the fallback usage seen in [the] \texttt{text} property, without adding unsupported claims". The DI summary "is penalized for verbosity and for including unsupported details not evidenced by the Function". The reference "is minimal but correct and concise". \\
\bottomrule
\end{tabular}
\caption{A complete judge evaluation (DeepSeek candidates, module context). The DA summary states the fallback role of the property, a fact evidenced only by the \texttt{text} property elsewhere in the module. The DI summary compensates with speculation the judge penalizes as unsupported. Candidate texts are abbreviated with [...] and available in full in the replication package.}
\label{fig:judge_outputs}
\Description{Example judge evaluation comparing the reference, documentation-aware, and documentation-independent candidate summaries for the apparent encoding property of the requests Response class.}
\end{figure}

\begin{table}[!t]
\centering
\caption{Win rates (\%) and Wilcoxon signed-rank test results by judged model and context ($n=984$--$990$ valid rankings per row, Claude Sonnet 4.6 judge). Highlighted p < .05.}
\label{tab:judge_win_rates}
\footnotesize
\setlength{\tabcolsep}{6pt}
\begin{tabular}{@{}l l r r r r r@{}}
\toprule
Model & Context & Ref & DA & DI & $p$ & r \\
\midrule
DeepSeek v3.2 & Class  & 42.4 & 31.0 & 26.6 & \textbf{<.0001} & 0.159 \\
DeepSeek v3.2 & Module & 47.1 & 27.5 & 25.4 & \textbf{.0162} & 0.077 \\
\addlinespace
gemini-3-flash-preview & Class  & 8.1 & 47.3 & 44.6 & \textbf{.0004} & 0.113 \\
gemini-3-flash-preview & Module & 9.6 & 47.6 & 42.8 & \textbf{.0006} & 0.109 \\
\addlinespace
gpt-4o-mini & Class  & 51.5 & 28.3 & 20.2 & \textbf{<.0001} & 0.134 \\
gpt-4o-mini & Module & 53.4 & 27.0 & 19.6 & \textbf{.0006} & 0.109 \\
\bottomrule
\end{tabular}
\end{table}

\begin{table}[!t]
\centering
\caption{Mann--Whitney U test for reference docstring length when reference is ranked first, by judged model and context. Values are reported as $n$, median$\pm$std. Highlighted p < .05.}
\label{tab:ref_rank1_mwu}
\footnotesize
\setlength{\tabcolsep}{5pt}
\begin{tabular}{@{}l l l r r@{}}
\toprule
Model & Context & Ref Rank1 (median$\pm$std) & $p$ & r \\
\midrule
GPT-4o-mini & Class  & $n=509,\ 157\pm365$ & \textbf{<.0001} & -0.284 \\
GPT-4o-mini & Module & $n=529,\ 157\pm357$ & \textbf{<.0001} & -0.288 \\
\addlinespace
gemini-3-flash-preview & Class  & $n=80,\ 316\pm508$ & \textbf{<.0001} & -0.207 \\
gemini-3-flash-preview & Module & $n=95,\ 257\pm464$ & \textbf{<.0001} & -0.194 \\
\addlinespace
DeepSeek v3.2 & Class  & $n=418,\ 162\pm354$ & \textbf{<.0001} & -0.250 \\
DeepSeek v3.2 & Module & $n=464,\ 166\pm364$ & \textbf{<.0001} & -0.281 \\
\bottomrule
\end{tabular}
\end{table}

For RQ3, all three models produce higher similarity scores with a broader context, but the magnitude varies substantially. Gemini shows the strongest response to hierarchical context, with medium effect sizes for Baseline vs. DA (r = 0.31--0.36). GPT-4o-mini produces mostly weak effects (r = 0.07--0.26), and DeepSeek produces mostly weak effects (r = 0.04--0.18). The judge evaluation confirms the DA preference for all three models (r = 0.08--0.16), and the models differ sharply in how their outputs fare against the reference: GPT-4o-mini and DeepSeek references win 42--53\% of comparisons while Gemini references win under 10\%. Model choice moderates both the size of the context benefit and the degree to which surrounding documentation contributes. The effect is model-dependent in magnitude, ranging from mostly weak (DeepSeek and GPT-4o-mini) to medium (Gemini).

The effects of DA extend to repositories with a lower representation in the sample (astropy, matplotlib, and sympy account for 69.2\% of the rows). On BERTScore, the DA median exceeds the DI median in 12 of 12 repositories for GPT-4o-mini and DeepSeek, and in 11 of 12 for Gemini (only xarray reverses, by 0.0011). Macro-averaging the per-repository DA--DI differences, which weights each repository equally, preserves the advantage for all three models (BERTScore +0.011 for GPT-4o-mini, +0.064 for Gemini, +0.017 for DeepSeek). The macro-average over the nine non-dominant repositories alone remains positive (+0.010, +0.049, +0.017). The replication package contains DA and DI results per repository.

\section{Discussion}
\label{sec:discussion}

\subsection{Comparison With The Literature}

Our results align with and quantify the observations of Makharev and Ivanov \cite{11028203}: DA prompts consistently outperform DI prompts on similarity metrics, class-level DA context yields the highest pooled values, and the magnitude of improvement is mixed, suggesting that models differ in their ability to exploit hierarchical context and surrounding documentation. Our findings also offer a new perspective on the work of Haldar and Hockenmaier \cite{haldar-hockenmaier-2024-analyzing}, who reported that LLMs often rely more on function names than the actual code structure. We observed that, with a sufficiently broad view, such as that provided by module-level context, the model can leverage the surrounding code structure and existing documentation to generate a summary closer to the developer's intent. This effect is strongest for Gemini, which shows the largest sensitivity to hierarchical context among the three models tested.

\subsection{Implications For Research And Practice}

$N$-gram metrics such as BLEU and ROUGE are sensitive to length differences between generated and reference text. By not enforcing a specific docstring style, we reduce this mismatch and allow the metrics to better reflect semantic alignment. We further triangulate with BERTScore and an LLM-as-a-Judge evaluation, following Crupi et al. \cite{crupi2025effectiveness}, who found low correlation between human judgment and $n$-gram metrics for code summarization.

The judge evaluation shows DA significantly preferred over DI for all three models, but with mostly weak effect sizes (r = 0.08--0.16), even for Gemini, whose metric effects are medium (r = 0.31--0.36). This gap between metric and judge effects calls for explanation. Copying does not provide one, since Gemini's DA outputs show the lowest lexical overlap with the surrounding documentation of the three models (Jaccard 0.046 DA vs.\ 0.042 DI). A more consistent explanation is stylistic alignment. Gemini adopts the documentation conventions of the surrounding scope, which raises similarity to references written under those same conventions by more than it raises the informational quality the judge rewards. Output style points the same way. Gemini produces the shortest summaries (median 312 characters at baseline), leaving the most headroom to absorb reference-like phrasing from context, while DeepSeek's verbose outputs (median 1,029) dilute overlap with any single source, consistent with its mostly weak effects.

Figure~\ref{fig:judge_outputs} illustrates what surrounding documentation contributes and what the judge rewards. The DA summary states the fallback role of the property, a fact evidenced only by the \texttt{text} property elsewhere in the module, and ranks first. The DI summary, generated without documentation cues, compensates with speculation about protocol headers and memory usage that the judge penalizes as unsupported, and the terse reference ranks second. The example matches the aggregate rationale statistics. The judge cites accuracy or correctness in 97\% of rationales, penalizes verbose or redundant summaries in 54\%, and praises detail or comprehensiveness in under 2\%. The DA advantage appears across the repositories and across both short and long summaries.

We also observe a weak but consistent dependence on reference comment length. As references become longer, they are increasingly ranked first by the judge, so improvements from context augmentation are most visible when the existing reference is under-specified rather than already comprehensive. For practitioners and tool developers, prompts should include broader hierarchical context and can benefit from retaining nearby documentation. Adding context increases average prompt size from 277 tokens for the Function-only baseline to 4,506 for Class DA and 13,233 for Module DA.

\subsection{Risks And Mitigation}
Generated docstrings can be plausible but incorrect (wrong parameter meanings, side effects, exceptions, or pre- and post-conditions). If adopted without review, such documentation may increase defects, maintenance costs, or security vulnerabilities by encouraging incorrect API usage. We therefore encourage human-in-the-loop review: treat generated docstrings as drafts requiring developer validation, especially for public APIs, security-sensitive code, or code with complex side effects.

\section{Threats To Validity}
\label{sec:threats_to_validity}
Drawing on Maxwell's worldview perspective \cite{maxwell1992understanding} and Petersen et al. \cite{petersen2013worldviews}, which include positivist aspects, we detail the key threats to validity identified in our study and the mitigation strategies considered.

\subsection{Theoretical Validity}
The metrics used in this study serve the key purpose of indicating developer intent. Although these metrics are widely used in the literature, $n$-gram metrics such as BLEU are sensitive to length differences between generated and reference docstrings. To mitigate this threat, we complement $n$-gram metrics with BERTScore for semantic similarity and an LLM-as-a-Judge evaluation for holistic quality assessment, supported by findings from Crupi et al. \cite{crupi2025effectiveness}.

A related concern is that the DA advantage reflects summary length rather than context use. Length correlates negatively with every metric (Spearman rho between -0.10 and -0.34), so longer summaries do not score higher, and DA still exceeds DI within every length quartile. The most concise model, Gemini, also shows the largest effect, so the result is not due to any single length-sensitive measure.

Because the SWE-Bench snapshot predates every model's knowledge cutoff, the entire sample may have been seen during pre-training. Our design controls for this. Each of the 6,932 functions appears in all five conditions, so memorized content contributes equally to every condition and cannot produce the condition differences our conclusions point to. Three observations argue against recall as the driver. Generated summaries are roughly six times longer than their references rather than reproductions of them. DA summaries are barely more lexically similar to the surrounding documentation than DI summaries. The DA advantage persists in the repositories least represented in our sample (Section~\ref{sec:judge_analysis}). A direct test would contrast functions committed after the models' knowledge cutoffs with pre-cutoff functions, which requires code newer than the pinned SWE-Bench snapshots. We leave it to future work.

Under DA, a model might paraphrase neighboring docstrings instead of reasoning over the code. On 400 sampled class-level targets per model, DA summaries are only marginally more similar to the surrounding documentation than DI summaries (Jaccard 0.065 vs.\ 0.058; DA higher in 59\% of pairs), and the shared tokens are mostly generic API vocabulary, so the DA improvement does not come from copying the surrounding documentation.

All judge-based results are based on a single judge model. A judge outside the generators' families avoids intra-family self-preference but has its own preferences. We mitigated this risk by running a blinded 20-task calibration (Section~\ref{evaluation}) which shows the judge agreeing with a human rater at the level the raters agree with each other, supporting the aggregate comparisons.

\subsection{External Generalizability}
The 12 repositories are Python projects, so our findings may not generalize to languages with different documentation conventions. Within the sample, three repositories (astropy, matplotlib, and sympy) contribute 69.2\% of the filtered functions. The repository-balanced analysis in Section~\ref{sec:judge_analysis} shows the DA advantage does not depend on this imbalance, but generalization beyond these domains remains untested. Our claims are likewise scoped to the three evaluated model families. The negligible-to-medium spread in effect sizes across them cautions against extrapolating to other families, including smaller open-weight and code-specialized models.

\subsection{Descriptive Validity}
Future retraining of the LLM with a current knowledge cutoff may affect experimental outcomes, as earlier revisions are not accessible via the current API. As an inherent consequence of working with closed LLMs offered as a service, this limitation undermines the reproducibility of results over time. To mitigate this threat, we employed the open-source model provider Deepseek in our analysis and noted the knowledge cutoff dates for each model, allowing for long-term replicability of results.

\section{Conclusion}
Generated docstrings in both Documentation-Independent (DI) and Documentation-Aware (DA) scenarios are longer than their references, and pooled BERTScore, BLEU, METEOR, and ROUGE values peak under class-level DA context. Kruskal--Wallis tests confirm that metric distributions differ across conditions for all three models, and Dunn--Bonferroni post-hoc tests show that most pairwise contrasts are statistically significant. The effect size varies by model, with Gemini showing medium effects for Baseline vs. DA (r = 0.31--0.36), GPT-4o-mini mostly weak effects (r = 0.07--0.26), and DeepSeek mostly weak effects (r = 0.04--0.18). The benefit of hierarchical context and surrounding documentation is consistent in direction but model-dependent in magnitude. Gemini produced summaries closest to the developer-written references, achieving the highest mean scores on all four metrics in every condition.

We triangulated our findings with a Claude Sonnet 4.6 LLM-as-a-Judge ranking on 1000 stratified samples per model using adequacy, conciseness, and fluency criteria. DA outputs are significantly preferred over DI for all three models, with mostly weak effect sizes. Longer developer-written references are more likely to be ranked first, suggesting that the headroom for LLM improvements depends on the completeness of the existing documentation.

Several directions remain for future work. The repositories used originate from SWE-bench~\cite{jimenez2024swebenchlanguagemodelsresolve} and are publicly available, so the evaluated models may have encountered them during pre-training. Comparing performance on functions present in training corpora against those absent, for example functions committed after March 2025, the most recent knowledge cutoff among the evaluated models, would separate contextual reasoning from memorization. This requires sampling commits newer than the pinned SWE-Bench snapshots. Repeating the study with smaller open-weight models (7B and 13B parameters) would clarify whether context sensitivity generalizes across model sizes~\cite{sun2024sourcecodesummarizationera}. Extending the hierarchy to repository-wide context~\cite{11028203, lomshakov2024proconsul} would test whether the modest effect sizes observed at the module level grow when cross-module dependencies enter the prompt. Applying the study to languages beyond Python would test generalizability across different documentation conventions and structural norms.

\section*{ARTIFACT AVAILABILITY}
Code, prompts and data supporting this study are openly available at: \newline \url{https://doi.org/10.6084/m9.figshare.31046290.v2}, under the MIT (code and prompts) and CC BY 4.0 (data) licenses.

\bibliographystyle{ACM-Reference-Format}
\bibliography{references}

@article{ZHOU2022106761,
title = {Summarizing source code with hierarchical code representation},
journal = {Information and Software Technology},
volume = {143},
pages = {106761},
year = {2022},
issn = {0950-5849},
doi = {https://doi.org/10.1016/j.infsof.2021.106761},
url = {https://www.sciencedirect.com/science/article/pii/S095058492100207X},
author = {Ziyi Zhou and Huiqun Yu and Guisheng Fan and Zijie Huang and Xingguang Yang}
}

@article{alon2018code2veclearningdistributedrepresentations,
author = {Alon, Uri and Zilberstein, Meital and Levy, Omer and Yahav, Eran},
title = {code2vec: learning distributed representations of code},
year = {2019},
issue_date = {January 2019},
publisher = {Association for Computing Machinery},
address = {New York, NY, USA},
volume = {3},
number = {POPL},
url = {https://doi.org/10.1145/3290353},
doi = {10.1145/3290353},
journal = {Proc. ACM Program. Lang.},
month = jan,
articleno = {40},
numpages = {29}
}

@inproceedings{petersen2013worldviews,
  title={Worldviews, Research Methods, and their Relationship to Validity in Empirical Software Engineering Research},
  author={Petersen, Kai and Gencel, Cigdem},
  booktitle={2013 Joint Conference of the 23rd International Workshop on Software Measurement (IWSM) and the Eighth International Conference on Software Process and Product Measurement (Mensura)},
  year={2013},
  note={Kai Petersen: School of Computing, Blekinge Institute of Technology, Karlskrona, Sweden; Cigdem Gencel: Faculty of Computer Science, Free University of Bolzano/Bozen, Italy},
  publisher = {IEEE Computer Society},
  address = {USA},
  pages = {81-89},
  numpages = {9}

}

@article{maxwell1992understanding,
  title={Understanding and validity in qualitative research},
  author={Maxwell, J. A.},
  journal={Harvard Educational Review},
  volume={62},
  number={3},
  pages={279--301},
  year={1992}
}

@inproceedings{ahmed2024automaticsemanticaugmentation,
  title     = {Automatic Semantic Augmentation of Language Model Prompts (for Code Summarization)},
  author    = {Ahmed, Toufique and Pai, Kunal Suresh and Devanbu, Premkumar and Barr, Earl T.},
  booktitle = {Proceedings of the 46th International Conference on Software Engineering},
  year      = {2024},
  doi       = {10.1145/3597503.3639183},
  url       = {https://dl.acm.org/doi/10.1145/3597503.3639183},
  publisher = {Association for Computing Machinery},
  address   = {New York, NY, USA},
  articleno = {220},
  numpages = {13},

}

@ARTICLE{XIA2094,
  author={Xia, Xin and Bao, Lingfeng and Lo, David and Xing, Zhenchang and Hassan, Ahmed E. and Li, Shanping},
  journal={IEEE Transactions on Software Engineering}, 
  title={Measuring Program Comprehension: A Large-Scale Field Study with Professionals}, 
  year={2017},
  month={07},
  volume={PP},
  pages={1-1},
  doi={10.1109/TSE.2017.2734091}}

@article{martinlopez2026morecode,
  author={Martin-Lopez, Alberto and Tufano, Rosalia and Guglielmi, Emanuela and S{\'a}nchez, Ana B and Sanz, Ana Lourdes and Scalabrino, Simone and Oliveto, Rocco and Segura, Sergio and Bavota, Gabriele},
  title={More Code, Less Understanding? On the Impact of AI Assistants on Developers' Productivity and Code Ownership},
  journal={IEEE Transactions on Software Engineering},
  year={2026}}

@article{kampenes2007systematic,
  title={A systematic review of effect size in software engineering experiments},
  author={Kampenes, Vigdis By and Dyb{\aa}, Tore and Hannay, Jo E and Sj{\o}berg, Dag IK},
  journal={Information and Software Technology},
  volume={49},
  number={11-12},
  pages={1073--1086},
  year={2007},
  publisher={Elsevier}
}

@software{spss_software_2025,
  title        = {{IBM SPSS Statistics}},
  author       = {{IBM Corp.}},
  year         = {2025},
  version      = {Version 29.0},
  publisher    = {IBM},
  url          = {https://www.ibm.com/products/spss-statistics}
}

@inproceedings{von-werra-etal-2022-evaluate,
    title = "Evaluate {\&} Evaluation on the Hub: Better Best Practices for Data and Model Measurements",
    author = "Von Werra, Leandro  and
      Tunstall, Lewis  and
      Thakur, Abhishek  and
      Luccioni, Sasha  and
      Thrush, Tristan  and
      Piktus, Aleksandra  and
      Marty, Felix  and
      Rajani, Nazneen  and
      Mustar, Victor  and
      Ngo, Helen",
    editor = "Che, Wanxiang  and
      Shutova, Ekaterina",
    booktitle = "Proceedings of the 2022 Conference on Empirical Methods in Natural Language Processing: System Demonstrations",
    month = dec,
    year = "2022",
    address = "Abu Dhabi, UAE",
    publisher = "Association for Computational Linguistics",
    url = "https://aclanthology.org/2022.emnlp-demos.13/",
    doi = "10.18653/v1/2022.emnlp-demos.13",
    pages = "128--136"
}

@inproceedings{sun2024sourcecodesummarizationera,
author = {Sun, Weisong and Miao, Yun and Li, Yuekang and Zhang, Hongyu and Fang, Chunrong and Liu, Yi and Deng, Gelei and Liu, Yang and Chen, Zhenyu},
title = {Source Code Summarization in the Era of Large Language Models},
year = {2025},
isbn = {9798331505691},
publisher = {IEEE Press},
url = {https://doi.org/10.1109/ICSE55347.2025.00034},
address = {USA},
doi = {10.1109/ICSE55347.2025.00034},
booktitle = {Proceedings of the IEEE/ACM 47th International Conference on Software Engineering},
pages = {1882-1894},
numpages = {13},
location = {Ottawa, Ontario, Canada},
series = {ICSE '25}
}

@inproceedings{qin2023chatgptgeneralpurposenaturallanguage,
    title = "Is {C}hat{GPT} a General-Purpose Natural Language Processing Task Solver?",
    author = "Qin, Chengwei  and
      Zhang, Aston  and
      Zhang, Zhuosheng  and
      Chen, Jiaao  and
      Yasunaga, Michihiro  and
      Yang, Diyi",
    editor = "Bouamor, Houda  and
      Pino, Juan  and
      Bali, Kalika",
    booktitle = "Proceedings of the 2023 Conference on Empirical Methods in Natural Language Processing",
    month = dec,
    year = "2023",
    address = "Singapore",
    publisher = "Association for Computational Linguistics",
    url = "https://aclanthology.org/2023.emnlp-main.85/",
    doi = "10.18653/v1/2023.emnlp-main.85",
    pages = "1339--1384"
}

@inproceedings{Papineni2002BLEU,
author = {Papineni, Kishore and Roukos, Salim and Ward, Todd and Zhu, Wei-Jing},
title = {BLEU: a method for automatic evaluation of machine translation},
year = {2002},
publisher = {Association for Computational Linguistics},
address = {USA},
url = {https://doi.org/10.3115/1073083.1073135},
doi = {10.3115/1073083.1073135},
booktitle = {Proceedings of the 40th Annual Meeting on Association for Computational Linguistics},
pages = {311-318},
numpages = {8},
location = {Philadelphia, Pennsylvania},
series = {ACL '02}
}

@inproceedings{Banerjee2005METEOR,
author = {Lavie, Alon and Agarwal, Abhaya},
title = {Meteor: an automatic metric for MT evaluation with high levels of correlation with human judgments},
year = {2007},
publisher = {Association for Computational Linguistics},
address = {USA},
booktitle = {Proceedings of the Second Workshop on Statistical Machine Translation},
pages = {228-231},
numpages = {4},
location = {Prague, Czech Republic},
series = {StatMT '07}
}

@inproceedings{Lin2004ROUGE,
    title = "{ROUGE}: A Package for Automatic Evaluation of Summaries",
    author = "Lin, Chin-Yew",
    booktitle = "Text Summarization Branches Out",
    month = jul,
    year = "2004",
    address = "Barcelona, Spain",
    publisher = "Association for Computational Linguistics",
    url = "https://aclanthology.org/W04-1013/",
    pages = "74--81"
}

@inproceedings{10.1145/1858996.1859006,
author = {Sridhara, Giriprasad and Hill, Emily and Muppaneni, Divya and Pollock, Lori and Vijay-Shanker, K.},
title = {Towards automatically generating summary comments for Java methods},
year = {2010},
isbn = {9781450301169},
publisher = {Association for Computing Machinery},
address = {New York, NY, USA},
url = {https://doi.org/10.1145/1858996.1859006},
doi = {10.1145/1858996.1859006},
booktitle = {Proceedings of the 25th IEEE/ACM International Conference on Automated Software Engineering},
pages = {43-52},
numpages = {10},
location = {Antwerp, Belgium},
series = {ASE '10}
}

@Article{sym14030471,
author = {Zhang, Chunyan and Wang, Junchao and Zhou, Qinglei and Xu, Ting and Tang, Ke and Gui, Hairen and Liu, Fudong},
title = {A Survey of Automatic Source Code Summarization},
journal = {Symmetry},
volume = {14},
year = {2022},
number = {3},
article-number = {471},
url = {https://www.mdpi.com/2073-8994/14/3/471},
issn = {2073-8994},
doi = {10.3390/sym14030471}
}

@inproceedings{feng2020codebertpretrainedmodelprogramming,
    title = "{C}ode{BERT}: A Pre-Trained Model for Programming and Natural Languages",
    author = "Feng, Zhangyin  and
      Guo, Daya  and
      Tang, Duyu  and
      Duan, Nan  and
      Feng, Xiaocheng  and
      Gong, Ming  and
      Shou, Linjun  and
      Qin, Bing  and
      Liu, Ting  and
      Jiang, Daxin  and
      Zhou, Ming",
    editor = "Cohn, Trevor  and
      He, Yulan  and
      Liu, Yang",
    booktitle = "Findings of the Association for Computational Linguistics: EMNLP 2020",
    month = nov,
    year = "2020",
    address = "Online",
    publisher = "Association for Computational Linguistics",
    url = "https://aclanthology.org/2020.findings-emnlp.139/",
    doi = "10.18653/v1/2020.findings-emnlp.139",
    pages = "1536--1547"
}

@misc{jimenez2024swebenchlanguagemodelsresolve,
      title={SWE-bench: Can Language Models Resolve Real-World GitHub Issues?}, 
      author={Carlos E. Jimenez and John Yang and Alexander Wettig and Shunyu Yao and Kexin Pei and Ofir Press and Karthik Narasimhan},
      year={2024},
      eprint={2310.06770},
      archivePrefix={arXiv},
      primaryClass={cs.CL},
      url={https://arxiv.org/abs/2310.06770}, 
}

@ARTICLE{crupi2025effectiveness,
  author={Crupi, Giuseppe and Tufano, Rosalia and Velasco, Alejandro and Mastropaolo, Antonio and Poshyvanyk, Denys and Bavota, Gabriele},
  journal={IEEE Transactions on Software Engineering}, 
  title={On the Effectiveness of LLM-as-a-Judge for Code Generation and Summarization}, 
  year={2025},
  volume={51},
  number={8},
  pages={2329-2345},
  doi={10.1109/TSE.2025.3586082}}

@inproceedings{10.5555/3666122.3668142,
author = {Zheng, Lianmin and Chiang, Wei-Lin and Sheng, Ying and Zhuang, Siyuan and Wu, Zhanghao and Zhuang, Yonghao and Lin, Zi and Li, Zhuohan and Li, Dacheng and Xing, Eric P. and Zhang, Hao and Gonzalez, Joseph E. and Stoica, Ion},
title = {Judging LLM-as-a-judge with MT-bench and Chatbot Arena},
year = {2023},
publisher = {Curran Associates Inc.},
address = {Red Hook, NY, USA},
booktitle = {Proceedings of the 37th International Conference on Neural Information Processing Systems},
articleno = {2020},
numpages = {29},
location = {New Orleans, LA, USA},
series = {NIPS '23}
}

@inproceedings{chen-etal-2025-beyond,
    title = "Beyond the Surface: Measuring Self-Preference in {LLM} Judgments",
    author = "Chen, Zhi-Yuan  and
      Wang, Hao  and
      Zhang, Xinyu  and
      Hu, Enrui  and
      Lin, Yankai",
    editor = "Christodoulopoulos, Christos  and
      Chakraborty, Tanmoy  and
      Rose, Carolyn  and
      Peng, Violet",
    booktitle = "Proceedings of the 2025 Conference on Empirical Methods in Natural Language Processing",
    month = nov,
    year = "2025",
    address = "Suzhou, China",
    publisher = "Association for Computational Linguistics",
    url = "https://aclanthology.org/2025.emnlp-main.86/",
    doi = "10.18653/v1/2025.emnlp-main.86",
    pages = "1653--1672",
    ISBN = "979-8-89176-332-6"
}

@misc{ralph2020empirical,
      title={Empirical Standards for Software Engineering Research}, 
      author={Paul Ralph and Nauman bin Ali and Sebastian Baltes and Domenico Bianculli and Jessica Diaz and Yvonne Dittrich and Neil Ernst and Michael Felderer and Robert Feldt and Antonio Filieri and Breno Bernard Nicolau de França and Carlo Alberto Furia and Greg Gay and Nicolas Gold and Daniel Graziotin and Pinjia He and Rashina Hoda and Natalia Juristo and Barbara Kitchenham and Valentina Lenarduzzi and Jorge Martínez and Jorge Melegati and Daniel Mendez and Tim Menzies and Jefferson Molleri and Dietmar Pfahl and Romain Robbes and Daniel Russo and Nyyti Saarimäki and Federica Sarro and Janet Siegmund and Diomidis Spinellis and Miroslaw Staron and Klaas Stol and Margaret-Anne Storey and Davide Taibi and Damian Tamburri and Marco Torchiano and Christoph Treude and Burak Turhan and Xiaofeng Wang and Sira Vegas}, 
      year={2020}, 
      eprint={2010.03525}, 
      archivePrefix={arXiv}, 
      primaryClass={cs.SE}, 
      url={https://arxiv.org/abs/2010.03525} }

@inproceedings{zhang2020bertscore,
  title={BERTScore: Evaluating Text Generation with BERT},
  author={Tianyi Zhang* and Varsha Kishore* and Felix Wu* and Kilian Q. Weinberger and Yoav Artzi},
  booktitle={International Conference on Learning Representations},
  year={2020},
  url={https://openreview.net/forum?id=SkeHuCVFDr}
}

@misc{baltes2025guidelinesempiricalstudiessoftware,
      title={Guidelines for Empirical Studies in Software Engineering involving Large Language Models}, 
      author={Sebastian Baltes and Florian Angermeir and Chetan Arora and Marvin Muñoz Barón and Chunyang Chen and Lukas Böhme and Fabio Calefato and Neil Ernst and Davide Falessi and Brian Fitzgerald and Davide Fucci and Marcos Kalinowski and Stefano Lambiase and Daniel Russo and Mircea Lungu and Lutz Prechelt and Paul Ralph and Rijnard van Tonder and Christoph Treude and Stefan Wagner},
      year={2025},
      eprint={2508.15503},
      archivePrefix={arXiv},
      primaryClass={cs.SE},
      url={https://arxiv.org/abs/2508.15503}, 
}

@inproceedings{haldar-hockenmaier-2024-analyzing,
    title = "Analyzing the Performance of Large Language Models on Code Summarization",
    author = "Haldar, Rajarshi  and
      Hockenmaier, Julia",
    editor = "Calzolari, Nicoletta  and
      Kan, Min-Yen  and
      Hoste, Veronique  and
      Lenci, Alessandro  and
      Sakti, Sakriani  and
      Xue, Nianwen",
    booktitle = "Proceedings of the 2024 Joint International Conference on Computational Linguistics, Language Resources and Evaluation (LREC-COLING 2024)",
    month = may,
    year = "2024",
    address = "Torino, Italia",
    publisher = "ELRA and ICCL",
    url = "https://aclanthology.org/2024.lrec-main.89/",
    pages = "995--1008"
}

@inproceedings{lomshakov2024proconsul,
    title     = {{ProConSuL}: Project Context for Code Summarization with {LLMs}},
    author    = {Lomshakov, Vadim and Podivilov, Andrey and Savin, Sergey and Baryshnikov, Oleg and Lisevych, Alena and Nikolenko, Sergey I.},
    booktitle = {Proceedings of the 2024 Conference on Empirical Methods in Natural Language Processing: Industry Track},
    year      = {2024},
    pages     = {866--880},
    publisher = {Association for Computational Linguistics},
    doi       = {10.18653/v1/2024.emnlp-industry.65},
    url       = {https://aclanthology.org/2024.emnlp-industry.65},
}

@INPROCEEDINGS {11028203,
author = { Makharev, Vladimir and Ivanov, Vladimir },
booktitle = { 2025 IEEE/ACM International Workshop on Large Language Models for Code (LLM4Code) },
title = {{ Code Summarization Beyond Function Level }},
year = {2025},
volume = {},
ISSN = {},
pages = {153-160},
doi = {10.1109/LLM4Code66737.2025.00024},
url = {https://doi.ieeecomputersociety.org/10.1109/LLM4Code66737.2025.00024},
publisher = {IEEE Computer Society},
address = {Los Alamitos, CA, USA},
month =May}

\end{document}